\documentclass[runningheads]{llncs}

\usepackage[T1]{fontenc}
\usepackage{tgtermes}
\usepackage{microtype}
\usepackage{amsmath}
\usepackage{float}
\usepackage{algorithm}
\usepackage{algpseudocode}
\usepackage{booktabs}
\usepackage{graphicx}
\usepackage{url}
\usepackage{bbding}
\usepackage{pifont}
\usepackage{tikz}
\usepackage{hyperref}
\usepackage{orcidlink}
\renewcommand{\orcidID}[1]{\textsuperscript{\orcidlink{#1}}}
\newcommand{\cn}[1]{\tikz[baseline=-0.5ex]\node[draw,circle,inner sep=0.3pt,minimum size=1.5ex]{\tiny\bfseries #1};}

\begin{document}


\title{RolloutPipe: Overlapping Pipelined Rollout and Training in Disaggregated On-Policy LLM Reinforcement Learning}
\titlerunning{RolloutPipe: Overlapping Pipelined Rollout and Training in Disaggregated LLM}

\author{Rongjian Chen\inst{1,2}\orcidID{0009-0009-1091-2752} \and
Jianmin Hu\inst{1,3}\orcidID{0009-0008-0083-3463} \and
Kejiang Ye\inst{1}\orcidID{0000-0001-6133-407X} \and
Minxian Xu\inst{1}\Envelope\orcidID{0000-0002-0046-5153} }
\authorrunning{R. Chen et al.}
\institute{Shenzhen Institutes of Advanced Technology, Chinese Academy of Sciences, Shenzhen, China \and
University of Chinese Academy of Sciences, Beijing, China
 \and
Southern University of Science and Technology, Shenzhen, China\\
\email{\{rj.chen2, jm.hu, kj.ye, mx.xu\}@siat.ac.cn}}

\maketitle

\begin{abstract}
Large language model (LLM) post-training for reasoning increasingly relies on reinforcement learning with verifiable rewards (RLVR),
where models learn from ground-truth feedback on mathematical, logical, and scientific tasks.
To enable flexible resource allocation and support heterogeneous training setups,
modern RLVR systems adopt disaggregated architectures that decouple rollout generation and policy training across independent GPU pools.
However, existing synchronous on-policy GRPO (Group Relative Policy Optimization) RLVR systems finish an entire rollout before starting training,
leaving the trainer GPU pool idle while rollout is still ongoing.
Asynchronous RL pipelines overlap the two stages, but at the cost of
training on stale data.
To address these challenges, we propose RolloutPipe, a post-training framework for disaggregated RLVR systems, which turns the fixed-weight rollout into a complete-group pipeline where
trainable groups move to the trainer while later groups are still being generated.
RolloutPipe achieves this through two techniques including complete-group pipelining (CGP) and frontier-group dispatch (FGD).
CGP dispatches each trainable complete group to the trainer FIFO as soon as group materialization finishes,
and FGD is an admission policy on the Rollout node that first admits requests for the frontier groups needed to form
the next training batch, so that trainer-ready groups arrive earlier and
more steadily.
The design starts training before the rollout completes while maintaining on-policy
correctness.
Evaluated on Qwen3-1.7B across four reasoning and science benchmarks and
twelve rollout settings, RolloutPipe shortens the rollout-to-train-end time
by 30.7\%--42.3\%,
and lowers the trainer waiting ratio by 37\%--76\% compared to Slime, a state-of-the-art rollout and training system.

\end{abstract}

\section{Introduction}

Post-training, fine-tuning a pre-trained model on task-specific data as distinct from initial pre-training on broad corpora,
has become the dominant way to improve LLM reasoning \cite{xu2026cloudnativedistributedsystemsefficient}.
It increasingly relies on RLVR,
where models learn from ground-truth feedback on mathematical, logical, and scientific tasks~\cite{deepseekr1}.
The central step in RLVR is {\em rollout}, which generates responses with the current model weights on the serving side.
Unlike inference on a fixed deployed model, rollout weights are updated between training rounds,
and each prompt is sampled multiple times so that a group of responses is ready when GRPO training begins.

To accommodate flexible resource allocation and heterogeneous setups,
modern RLVR systems adopt disaggregated architectures that decouple rollout generation and policy training across independent GPU pools.
Such generation-then-learning iteration echoes the cognitive computing loop of perceiving, reasoning, and self-updating:
a large language model acts as a cognitive engine that perceives tasks, produces reasoning traces, and receives verifiable feedback,
and the disaggregated deployment must balance generation and training so that newly perceived experience feeds learning with minimal latency.

However, existing disaggregated on-policy RL frameworks generally follow a serial rollout-then-train paradigm.
The training side receives data only after the entire rollout finishes,
leaving the training GPU pool idle throughout the rollout stage.
Some prompt groups have already become valid GRPO training units before
the full rollout completes, yet the serial path keeps them outside the
trainer FIFO, so the training GPUs sit idle even when runnable work already exists.
Existing RLVR systems provide rollout, reward, training, and
synchronization infrastructure~\cite{hybridflow,openrlhf,rlhfuse}, and 
asynchronous rollout, generation pipelining, and multi-weight pipelines
reduce this idleness by overlapping stages, but at the cost of stale data or
multiple weight snapshots~\cite{areal}.

To address these challenges without compromising on-policy correctness,
we propose \textit{RolloutPipe}, which turns an ongoing rollout into a
rollout-training pipeline while keeping the rollout weights of that round fixed.
The pipeline is built at the complete-group boundary, therefore, 
once a group becomes trainable it leaves the active rollout and enters training,
while the remaining groups continue consuming rollout resources.
Our key \textbf{contributions} are as follows.

\begin{itemize}
\item We identify the limitations of native Slime~\cite{slime} in disaggregated on-policy RL training. The serial execution of rollout and trainer leaves the training GPU pool idle throughout the entire rollout, and early-completed trainable complete groups are blocked outside the trainer FIFO by the rollout-completion barrier and cannot enter training in time.

\item We present RolloutPipe, a post-training framework built on native Slime that breaks the serial rollout-then-train barrier in disaggregated on-policy RL by turning an ongoing rollout into a rollout-training pipeline.

\item We design two mechanisms, complete-group pipelining (CGP) and frontier-group dispatch (FGD). CGP delivers each trainable complete group to the trainer FIFO as soon as it is ready, starting training before the rollout completes while preserving on-policy correctness. FGD is a Rollout-node admission policy that prioritizes requests for the \(F_{w}\) frontier groups forming the next trainer batch, so that trainable groups arrive earlier and more steadily.

\item We conduct end-to-end experiments on Qwen3-1.7B across four reasoning and scientific workloads, covering twelve settings under Slime, CGP, and CGP+FGD configurations. Compared with native Slime, RolloutPipe shortens the rollout-to-train-end main time by 30.7\%--42.3\% and lowers the trainer waiting ratio by 37\%--76\%.
\end{itemize}

\section{Background and Motivation}\label{sec:bg}

\subsection{Group-Level Constraint and Serial Idle}

Proximal Policy Optimization (PPO)-style Reinforcement Learning from Human Feedback (RLHF) organizes advantages and rewards per sample or trajectory~\cite{instructgpt}.
Its clipped surrogate objective relies on per-sample advantages and probability ratios between the current and old policies.
A single completed response is thus a valid training unit, enabling sample-level pipelines.

GRPO removes the learned critic and instead normalizes rewards group-wise.
For \(K\) responses under the same prompt, the advantage of each response is computed by subtracting the group mean reward and dividing by the group standard deviation.
The key structural difference from PPO is that the group statistics (mean and standard deviation) depend on all \(K\) responses, so no response is trainable until the entire group finishes reward and verifier computation, group-statistics computation, and equipping each sample with its advantage and loss mask.
We refer to this process as \emph{group materialization}.
The trainable unit is therefore the \emph{complete group}, not an individual sample.

\begin{figure}[!htbp]
\centering
\vspace{-0.5em}
\includegraphics[width=0.98\textwidth,trim=0 630 0 17,clip]{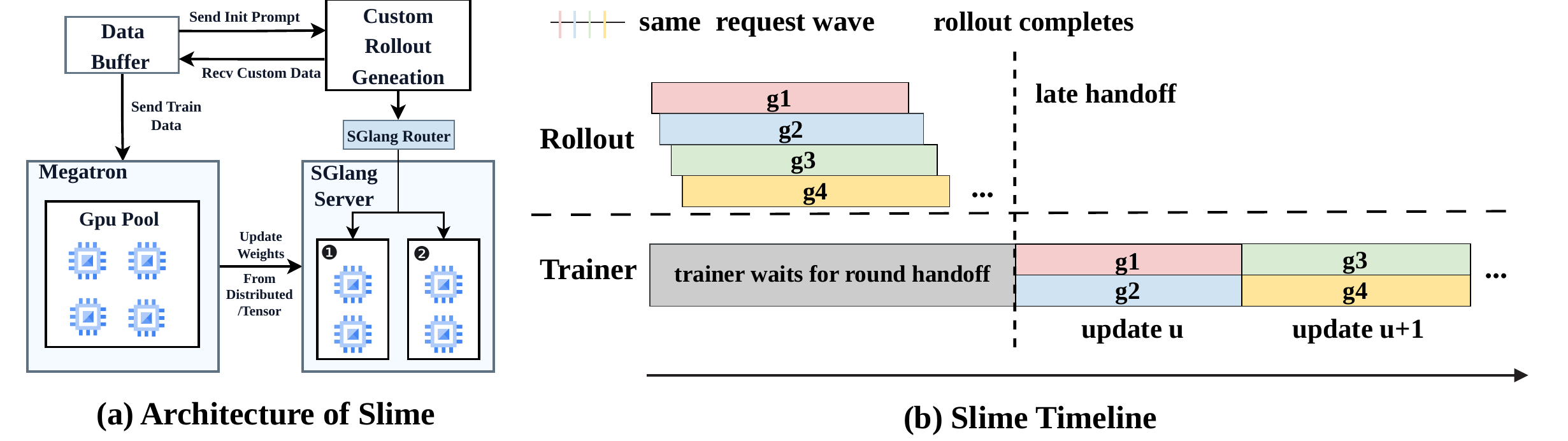}
\vspace{-1em}
\caption{(a) Slime's disaggregated architecture. (b) Serial rollout-then-train mode: early-completed groups wait outside the trainer FIFO, leaving it idle.}
\label{fig:baseline-timeline}
\vspace{-0.5em}
\end{figure}

This group-level dependency means GRPO cannot start training as soon as a single sample completes, unlike PPO.
However, existing synchronous systems (e.g., Slime, as shown in Figure~\ref{fig:baseline-timeline}(a)) respond by waiting for \emph{all} \(R\) groups
to finish before submitting the entire round to the trainer,
wasting the gap between early group completions and rollout completion.

As shown in Figure~\ref{fig:baseline-timeline}(b), in a rollout of \(R\) groups, groups do not complete simultaneously.
The time from the first group's materialization to the last (the \emph{staggered-completion window})
is idle trainer time that a synchronous system cannot recover.
Consider Slime on Qwen3-1.7B with the LSAT-AR workload, e.g. 
under the configuration \(R=96\) and \(K=8\),
the first group finishes at 200s after rollout start and the last finishes at 400s.
Then Slime submits all 96 groups to the trainer at the 400s.
Among the first 95 groups, the earliest one remains waiting outside the trainer FIFO for 200s,
during which it is already a valid training unit but the trainer remains idle.

For frameworks like Slime, this problem is especially severe when rollout and training resources are separated (Figure~\ref{fig:baseline-timeline}(a)).
Serving GPUs handle generation and reward computation, while training GPUs handle gradient updates,
and the two operate independently.
When the Rollout node has trainable complete groups ready, the training GPUs remain entirely idle
yet cannot use those groups. The serial rollout-then-train barrier keeps them outside the trainer.
In the evaluation configurations, Slime's end-to-end time has a trainer waiting ratio of 47\%--52\%.

\textit{\textbf{Observation 1.} The serial boundary of synchronous GRPO leaves the trainer idle for nearly half of the end-to-end time, even though early-completing groups are already valid training units.
Implication. We need a mechanism that lets complete groups enter training as soon as they finish materialization. This is the goal of our proposed CGP.}

\subsection{Unordered Completion and Frontier-Group Supply}

Even with complete groups entering the training side earlier,
pipeline utilization remains entirely dependent on the SGLang server schedule in Figure~\ref{fig:baseline-timeline}(a).
Standard SGLang admission is request-level FIFO, which operates on individual requests and is unaware of group membership.
Since all requests in a rollout arrive nearly simultaneously,
FIFO admission offers no control over which groups complete first.
Response lengths vary across prompts, rollout queues contend for GPU capacity, and verifier latency differs per task.
Because rollout capacity is evenly spread across all in-flight requests,
some groups that could finish earlier and reach the training side sooner
end up progressing at the same pace as non-frontier groups,
with no group-aware policy to prioritize them as training data.

This unordered completion directly affects the \emph{frontier-group arrival gap}.
Frontier groups are those whose completion will form the next training batch.
Suppose the training side consumes \(U=2\) groups per logical update, so
the trainer must wait for the first 2 trainable groups to arrive before starting the first update.
If group 0 finishes at 210s and group 1 finishes at 250s, the trainer waits 40s to form a batch.
During those 40s, group 0 is already in the trainer FIFO but cannot start training without group 1.
Subsequent \(U\)-groups face similar arrival gaps.

The root cause of large frontier-group arrival gaps is that default FIFO scheduling is oblivious to group boundaries.
Since each request receives the same scheduling opportunity,
frontier-group completion order depends on the arbitrary combination of prompt length,
response length, and verifier latency,
with no guarantee that frontier groups arrive earlier or more steadily.

\textit{\textbf{Observation 2. }Default FIFO admission is oblivious to group boundaries, where frontier and non-frontier groups progress at the same rate, so frontier-group completion order is determined by random factors.
Implication. We need a Rollout-node scheduling policy that prioritizes the frontier groups that will form the next training batch and makes them complete earlier and more steadily. This is the goal of our proposed FGD.}

\section{RolloutPipe Design}

\subsection{Overview}

\begin{figure}[!htbp]
\centering
\includegraphics[width=0.98\textwidth,trim=0 75 0 10,clip]{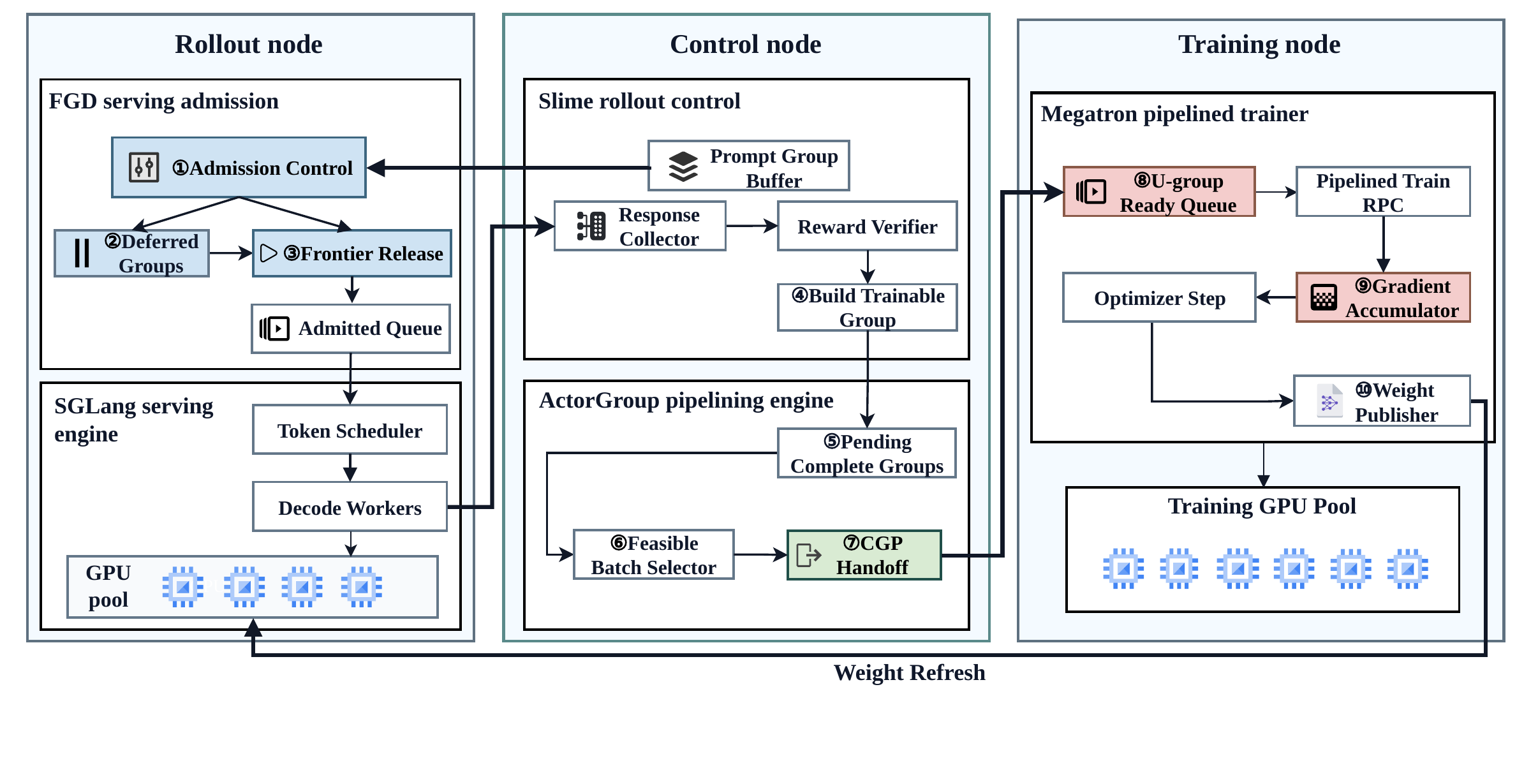}
\caption{RolloutPipe system architecture.
The two insertion points are FGD (serving-admission path, between rollout worker and SGLang router)
and CGP (complete-group handoff path, after group materialization).
The weight-refresh path (dashed) runs after the trainer drains all logical updates of the current rollout.}
\label{fig:rolloutpipe-architecture}
\end{figure}

RolloutPipe designs two mechanisms to break the serial rollout-then-train barrier,
as illustrated in Figure~\ref{fig:rolloutpipe-architecture}.
FGD is the Rollout-node optimization, where \cn{1} Admission Control maintains the frontier \(\mathcal{F}\) in the SGLang admission path,
admitting requests whose group is in \(\mathcal{F}\) through \cn{3} Frontier Release into the Admitted Queue,
while the rest are held in \cn{2} Deferred Groups, thereby reshaping the group completion order (Section~3.3).
CGP is the training-side optimization, where each group, after \cn{4} Build Trainable Group completes materialization,
enters \cn{5} Pending Complete Groups,
and after \cn{6} Feasible Batch Selector picks the group prefix, \cn{7} CGP Handoff forwards the batch to \cn{8} U-group Ready Queue,
enabling training to start before the rollout completes (Section~3.2).
On the training side, \cn{9} Gradient Accumulator tracks the number of consumed groups,
and \cn{10} Weight Publisher exports refreshed weights after all groups are consumed.
Together, these two mechanisms enable training to start before the rollout completes
while preserving on-policy correctness.
All groups in a round share the same fixed rollout weights,
the optimizer-step and weight-publish boundaries follow Slime,
and weight refresh runs only after the trainer drains all logical updates of the current rollout.

\begin{figure}[!htbp]
\centering
\includegraphics[width=0.9\textwidth,trim=0 175 0 18,clip]{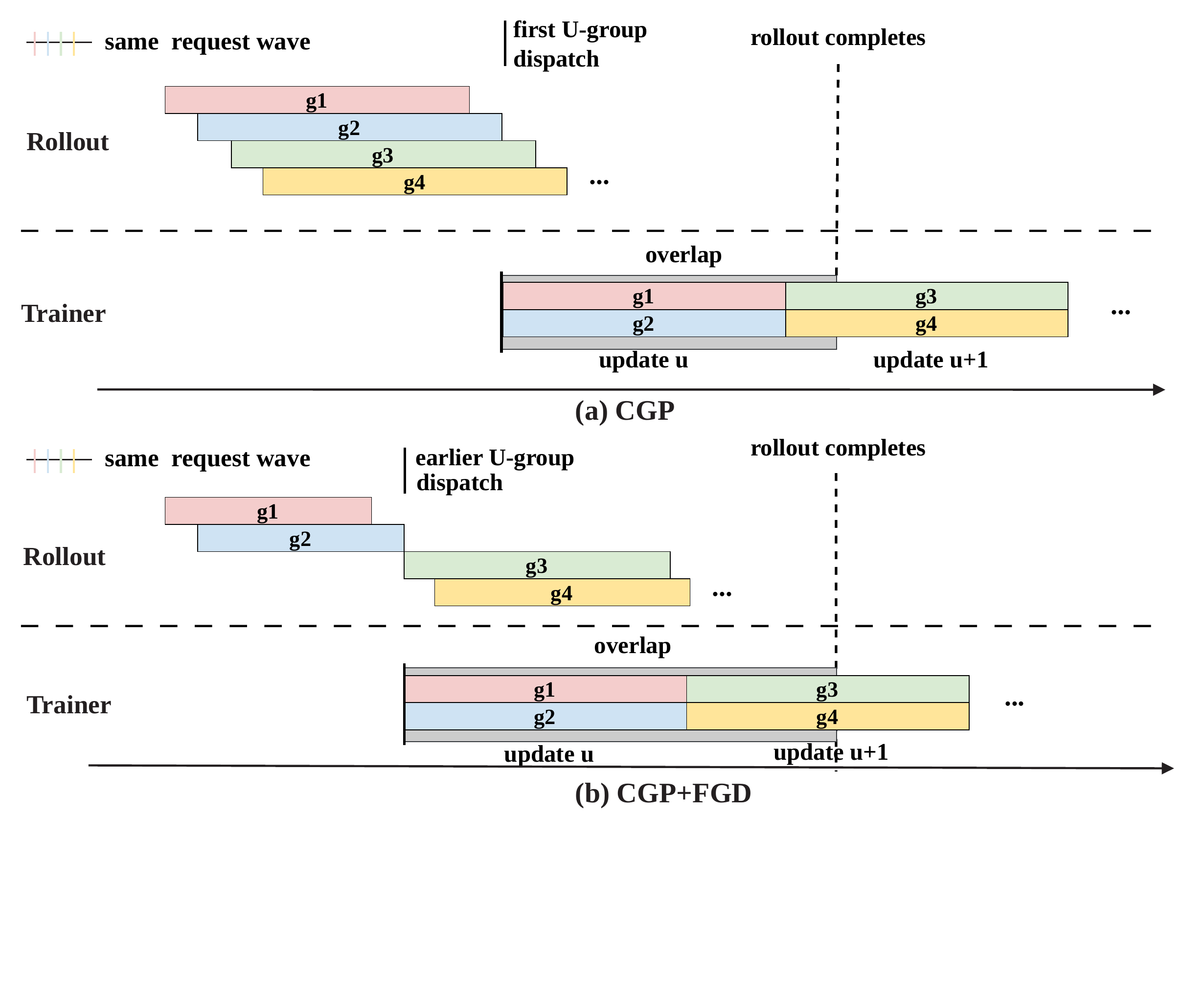}
\caption{Two illustrative RolloutPipe timelines.
(a)~CGP delivers each group to the trainer FIFO as soon as it materializes, so the first \(U\)-group dispatch occurs before the rollout completes.
(b)~CGP+FGD applies frontier-first admission, making the first dispatch arrive earlier.}
\label{fig:rolloutpipe-timeline}
\end{figure}

As illustrated in Figure~\ref{fig:rolloutpipe-timeline}, CGP delivers trainable complete groups to the trainer FIFO as soon as each group materializes, so the first \(U\)-group dispatch occurs before the rollout completes and overlaps with later groups' rollout.
CGP+FGD further applies frontier-first admission so that the frontier groups arrive earlier and the first \(U\)-group dispatch moves even earlier.
The Slime baseline, which submits all groups only after the rollout completes, is shown separately in Figure~\ref{fig:baseline-timeline}(b).
In essence, RolloutPipe refines the rollout-trainer synchronization granularity from the entire round to the complete-group level, turning Slime's idle gap into rollout-training overlap while preserving on-policy correctness.

\subsection{Training-Side Complete-Group Pipelining}

GRPO removes the learned critic and normalizes rewards group-wise. For \(K\) responses under the same prompt with rewards \(r_1,\ldots,r_K\), the advantage of response \(i\) is given by
\begin{equation}
A_i = \frac{r_i - \mu_{\text{group}}}{\sigma_{\text{group}} + \epsilon_{\mathrm{std}}},
\end{equation}
where \(\mu_{\text{group}}\) and \(\sigma_{\text{group}}\) are the mean and standard deviation computed from all \(K\) rewards in the group, and \(\epsilon_{\mathrm{std}}\) is a small stabilizer.
Because \(\mu_{\text{group}}\) and \(\sigma_{\text{group}}\) depend on all \(K\) responses, no response is trainable until its entire group finishes materialization. This is the constraint that CGP targets.
Let \(B\) denote the sample-level global batch size (total number of samples per training update).
then one logical update consumes \(U=B/K\) complete groups.

\textbf{CGP mechanism.}
Slime hands all \(R\) groups to the trainer when the rollout completes.
CGP instead appends each group to a FIFO maintained by the ActorGroup (the Slime component that wraps the trainer Ray actors and bridges rollout output to the trainer) as soon as it finishes materialization,
and once the FIFO accumulates \(U\) trainable groups,
the earliest \(U\) are dispatched to the trainer.

Let \(t_{\text{first}}^{(U)}\) denote the time at which the first \(U\) groups have all finished materialization
(i.e., the moment the first trainable batch is ready),
and \(t_{\text{complete}}\) the rollout-completion time.
The trainer-start time of the first update under each mode is given by
\begin{equation}\label{eq:cgp-start}
t_{\text{start}}^{\text{Slime}} = t_{\text{complete}}, \qquad
t_{\text{start}}^{\text{CGP}} = t_{\text{first}}^{(U)}.
\end{equation}
Since the first \(U\) groups usually finish long before the rollout completes,
CGP turns the trainer's idle wait from \(t_{\text{first}}^{(U)}\) to \(t_{\text{complete}}\) in Slime
into training time, where the recovery time is given by
\begin{equation}
\Delta t_{\text{CGP}} = t_{\text{complete}} - t_{\text{first}}^{(U)}.
\end{equation}
We define the trainer waiting ratio as the fraction of the end-to-end time in which the trainer is idle waiting for rollout data.
\begin{equation}
w_{\text{train}} = \frac{t_{\text{start}} - t_{\text{rollout\_start}}}{t_{\text{train\_end}} - t_{\text{rollout\_start}}}.
\end{equation}
Here \(t_{\text{start}}\) is the trainer start time, and \(t_{\text{rollout\_start}}\) and \(t_{\text{train\_end}}\) are the rollout start and training end times.
In Slime, training begins only after the rollout completes, so \(w_{\text{train}}\) is high.
CGP lowers \(w_{\text{train}}\) by moving \(t_{\text{start}}\) earlier and turning part of the idle wait into training time.
Section~\ref{sec:eval} quantifies this empirically.

Once the FIFO head accumulates \(U\) trainable complete groups,
Megatron executes forward, backward, and gradient accumulation.
The gradient accumulator tracks how many groups have been consumed.
the optimizer step triggers only when accumulation reaches \(U\),
preserving the same logical update semantics as the serial baseline.
In Slime, this entire pipeline starts only after the rollout completes, while CGP starts as soon as the first \(U\) groups materialize,
overlapping gradient computation with later groups' rollout execution.

\subsection{Frontier-Group Dispatch}

CGP enables training to start early but does not control the order in which groups complete.
In native SGLang, all \(R \times K\) requests are issued simultaneously and admitted via request-level FIFO, so groups complete with large inter-completion gaps and the trainer must wait until \(t_{\text{first}}^{(U)}\) before forming the first logical update.

\textbf{FGD mechanism.}
FGD keeps a frontier \(\mathcal{F}\) of at most \(F_{w}\) groups currently in service (\(|\mathcal{F}|\le F_{w}\)).
At each scheduling cycle, while \(|\mathcal{F}|<F_{w}\), the lowest-order group among the deferred groups is admitted and added to \(\mathcal{F}\). Requests whose group is not in \(\mathcal{F}\) are deferred.
When \(g\) completes \(K\) requests it leaves \(\mathcal{F}\), freeing a slot that the next cycle refills with the next lowest-order group.

Formally, let \(O=\{g_1,\ldots,g_R\}\) be the set of groups arrived but not yet serving-complete, and let \(\mathrm{order}(g)\) return the submission order of group \(g\) (smaller = earlier).
The frontier set is the \(F_{w}\) groups of lowest order in \(O\), defined as
\begin{equation}
\mathcal{F} = \arg\min^{\,F_{w}}_{g\in O}\, \mathrm{order}(g),
\end{equation}
where \(\arg\min^{F_{w}}\) selects the \(F_{w}\) groups with the smallest order values.
A request \(q\) is admitted into the rollout engine if and only if its group belongs to the frontier.
\begin{equation}
\mathrm{admit}(q) \iff g(q) \in \mathcal{F}, \qquad \text{otherwise } q \text{ enters Deferred}.
\end{equation}
Because \(|\mathcal{F}|=F_{w}\) bounds the number of groups served concurrently, FGD concentrates rollout capacity on the earliest groups, so the first \(U\) groups finish earlier and more steadily, thus lowering \(t_{\text{first}}^{(U)}\) in Eq.~\eqref{eq:cgp-start}.
\textbf{Example.} \(F_{w} = 2\), groups 1--6 arrived, \(\mathcal{F} = \{1, 2\}\).
After group 1 completes its \(K\) requests, \(\mathcal{F}\) updates to \(\{2, 3\}\).
For the training side, when \(|Q^{(r)}| \ge U = 2\), the trainer starts while groups 3--6 are still in rollout.

\textbf{Completion signals.}
FGD marks a group \(g\) as serving-complete once all \(K\) requests finish, and \(g\) is removed from \(\mathcal{F}\) and the candidate set.
CGP marks a group as trainable-complete after reward, verifier, and materialization, and appends it to the ActorGroup FIFO, forwarding the feasible prefix to the ready queue.

\section{Algorithm Design}

Algorithm~1 summarizes the combined FGD and CGP flow across the four components involved in RolloutPipe, namely FGD Admit, Rollout Worker, CGP Handoff, and Train.

{\setlength{\intextsep}{12pt}%
\begin{algorithm}[H]
\caption{RolloutPipe pipeline.}
\begin{algorithmic}[1]
\setlength{\itemsep}{0pt}%
\Statex \textbf{Input:} requests \(\{q\}\) grouped by \(g\); frontier width \(F_{w}\); update width \(U\).
\Statex \textbf{Output:} updated policy weights after consuming all \(R\) groups of the current rollout.
\Function{RolloutPipe}{}
    \State \(\mathbf{while}\) training not done:
    \State \(\quad\) \(\mathrm{FGD\_Admit}(\{q\}, F_{w})\)
    \State \(\quad\) \(\mathrm{Rollout\_Worker}()\)
    \State \(\quad\) \(\mathrm{CGP\_Handoff}(Q, U)\)
    \State \(\quad\) \(\mathrm{Train}(S_u)\)
\EndFunction
\Function{FGD\_Admit}{$\{q\}, F_{w}$}
    \State \(\mathbf{on}\) \(q\) arrives: admit if \(g(q)\in\mathcal{F}\), else \(q\!\to\!\)Deferred
    \State \(\mathbf{while}\;|\mathcal{F}|<F_{w}\): \(g^{*}\!\gets\!\arg\min_{g'\in\mathrm{Deferred}}\mathrm{order}(g')\); \(\mathcal{F}\!\gets\!\mathcal{F}\cup\{g^{*}\}\)
    \State \(\mathbf{when}\) \(g\) done: \(\mathcal{F}\!\gets\!\mathcal{F}\setminus\{g\}\)
\EndFunction
\Function{Rollout\_Worker}{}
    \State \(\mathbf{on}\) \(g\) complete: \(\mathrm{Materialize}(g)\to\) Pending Complete Groups
\EndFunction
\Function{CGP\_Handoff}{$Q, U$}
    \State \(b\gets\mathrm{FeasibleBatch}(Q)\); pop \(b\) to U-group Ready Queue
    \State \(\mathbf{if}\;|Q^{(r)}|\ge U\): \(S_u\gets Q^{(r)}[1{:}U]\to\)trainer
\EndFunction
\Function{Train}{$S_u$}
    \State consume \(S_u\) via Pipelined Train RPC, Grad Accumulator
    \State \(\mathbf{if}\) consumed\(=U\): Opt Step; \(\mathbf{if}\) all \(R\) consumed: Weight Publisher
\EndFunction
\end{algorithmic}
\end{algorithm}}

We implement RolloutPipe with Slime, Megatron-LM, SGLang, and Ray~\cite{slime,megatronlm,sglang,ray}.
Although the four functions are listed sequentially, they sit on three separate nodes and are event-driven.
The Rollout node advances admission and generation on request arrival and group completion,
the Control node hands off a group once it materializes,
and the training node starts an update whenever the ready queue accumulates \(U\) groups.
Rollout, materialization, handoff, and training therefore need not wait for one another but overlap along complete-group boundaries.
As shown in Figure~\ref{fig:rolloutpipe-architecture}, the three nodes carry the following responsibilities.

\textbf{Rollout node.}
This node handles frontier admission and generation (lines 8--15).
Prompt Group Buffer feeds the prompt groups of the current round into Admission Control,
which maintains the frontier \(\mathcal{F}\) (\(|\mathcal{F}|\le F_{w}\)), corresponding to the frontier-admission step in Algorithm~1.
When a request arrives, if its group is already in \(\mathcal{F}\),
Frontier Release admits it into the Admitted Queue.
Otherwise it is held in Deferred Groups.
At each scheduling cycle, while \(|\mathcal{F}|<F_{w}\),
the lowest-order group in Deferred Groups is admitted and added to \(\mathcal{F}\),
so the engine serves at most \(F_{w}\) groups at a time
and concentrates capacity on the earliest submitted groups.
Requests in the Admitted Queue are then scheduled by the SGLang serving engine.
The Token Scheduler allocates memory and compute,
Decode Workers run decoding on the GPU pool,
and the whole process generates responses for all requests of groups in \(\mathcal{F}\)
under the fixed rollout weights and returns completion notifications.
Once a group completes all \(K\) responses, it leaves \(\mathcal{F}\),
and the freed slot is refilled by the next cycle with the next lowest-order group.

\textbf{Control node.}
This node builds and hands off complete groups (lines 16--19), receiving the responses returned by the Rollout node.
Slime rollout control first builds each trainable group as follows.
The Response Collector gathers all responses of a group,
the Reward Verifier computes the reward and writes back the verifier result,
and Build Trainable Group then completes group materialization,
i.e., sample-record conversion, dynamic filtering, and group-statistics completion,
turning the group into a valid training unit.
The trainable group is then appended to Pending Complete Groups in the ActorGroup pipelining engine.
The Feasible Batch Selector picks the largest prefix of \emph{groups}
(counted in whole groups, never single requests)
that fits the per-GPU token budget. Even when a single group exceeds the budget,
it is still taken alone and adapted through dynamic micro-batching.
CGP Handoff then forwards this batch of groups to the U-group Ready Queue,
where it awaits consumption by the training node.

\textbf{Training node.}
This node performs gradient updates and weight publishing (lines 20--22), taking trainable groups from the U-group Ready Queue
and running forward and backward computation on the Training GPU Pool
through the Pipelined Train RPC,
feeding the results into the Gradient Accumulator.
Every \(U\) consumed groups trigger one Optimizer Step,
matching the logical-update boundary of the serial baseline.
Once all \(R\) groups of the current rollout are consumed,
the Weight Publisher exports the refreshed weights
to the SGLang serving engine (Weight Refresh) for use in the next rollout.

\textbf{Algorithm Complexity Analysis.}
Let \(R\) denote the number of prompt groups per rollout and \(K\) the number of sampled responses per prompt.
FGD and CGP introduce a control-flow overhead of \(\mathcal{O}(R \cdot K)\).
The per-rollout trainer-side forward and backward cost is \(\mathcal{O}(R \cdot K \cdot L^{2} \cdot d)\), identical to Slime,
since both process the same \(R \cdot K\) samples under the same Megatron microbatching,
where \(L\) is the sequence length and \(d\) is the hidden dimension.
Space overhead is also linear.
\textbf{Communication overhead.}
Group transfers incur network cost proportional to \(R\), but this overlaps with rollout and training, so communication overhead is hidden.

\section{Performance Evaluation}\label{sec:eval}

\subsection{Experimental Setup}

All experiments use Qwen3-1.7B as the policy backbone,
initialized from the same base checkpoint across Slime, CGP, and CGP+FGD.
It is the smallest model in the Qwen3 family with reasoning capability on our benchmarks,
enabling experiments across four workloads and varying R configurations under our GPU budget.
The training side employs \(8 \times\) RTX 4090 24GB GPUs with TP=4, DP=2,
and the Rollout node employs \(2 \times\) A100 40GB PCIe GPUs with TP=2.
The main results span four reasoning and scientific workloads,
each evaluated at \(R=32,64,96\) across three configurations over four rounds.
\begin{itemize}
  \item \textbf{LSAT-AR}~\cite{agieval}: LSAT analytical-reasoning
        problems from AGIEval and AR-LSAT, asking for valid arrangements under constraints.
  \item \textbf{Sci-XW}~\cite{scibench}: college-level
        math, chemistry, and physics problems from SciBench.
  \item \textbf{Sci-JL}~\cite{scibench}: a SciBench
        subset with a different problem distribution than Sci-XW.
  \item \textbf{OlyPhys}~\cite{olympiadbench}: math and
        physics olympiad problems from international and Chinese competitions.
\end{itemize}

The GRPO parameters are fixed at \(K=8\) and \(B=16\), giving \(U=2\) complete groups per logical update.
Here, \(R\) denotes the number of prompt groups in each rollout.
For FGD, the frontier width is set to \(F_{w}=U=2\), so that each batch of groups produced by the rollout side is exactly sufficient for one training update. Both \(F_{w}\) and \(U\) are configured based on the available GPU memory.
All configurations share the same generation and training hyperparameters,
with an exact-match accuracy reward and verifier.
Each result point reports the mean of four rounds per configuration.
Error bars show the sample standard deviation.
RolloutPipe's two mechanisms—CGP and FGD—are scheduling policies at the request-admission and complete-group level,
independent of model architecture,
so the qualitative findings are not specific to this model size.
Since CGP and FGD operate at the scheduling level and do not depend on model-internal parameters (such as hidden dimension or layer count), we expect these mechanisms to generalize to larger model scales.

We use three widely-used metrics as below:
\begin{itemize}
  \item \textbf{rollout-to-train-end}: the main time from rollout start to train end.
  \item \textbf{dispatch timing}: the time at which the first \(U\)-group logical batch is dispatched from the ready FIFO to the trainer.
  \item \textbf{training load}: response length, trainer compute time, and waiting ratio.
        Waiting ratio is trainer-side \(\mathrm{wait}/(\mathrm{wait}+\mathrm{compute})\).
        (trainer-side \(\mathrm{wait}/(\mathrm{wait}+\mathrm{compute})\)).
\end{itemize}

We mainly compare the state-of-the-art approach, \textbf{Slime}~\cite{slime} that uses the serial rollout and training way as baseline with our proposed approaches incorporated in RolloutPipe: 

\textbf{CGP} enables complete-group events, the training-side ready FIFO, and early gradient accumulation,
while using default request admission.

\textbf{CGP+FGD} adds the group-frontier admission controller under the same readiness predicate and training-side consumption rules.

\subsection{Results Analysis}

\begin{figure}[!htbp]
\centering
\includegraphics[width=0.7\textwidth]{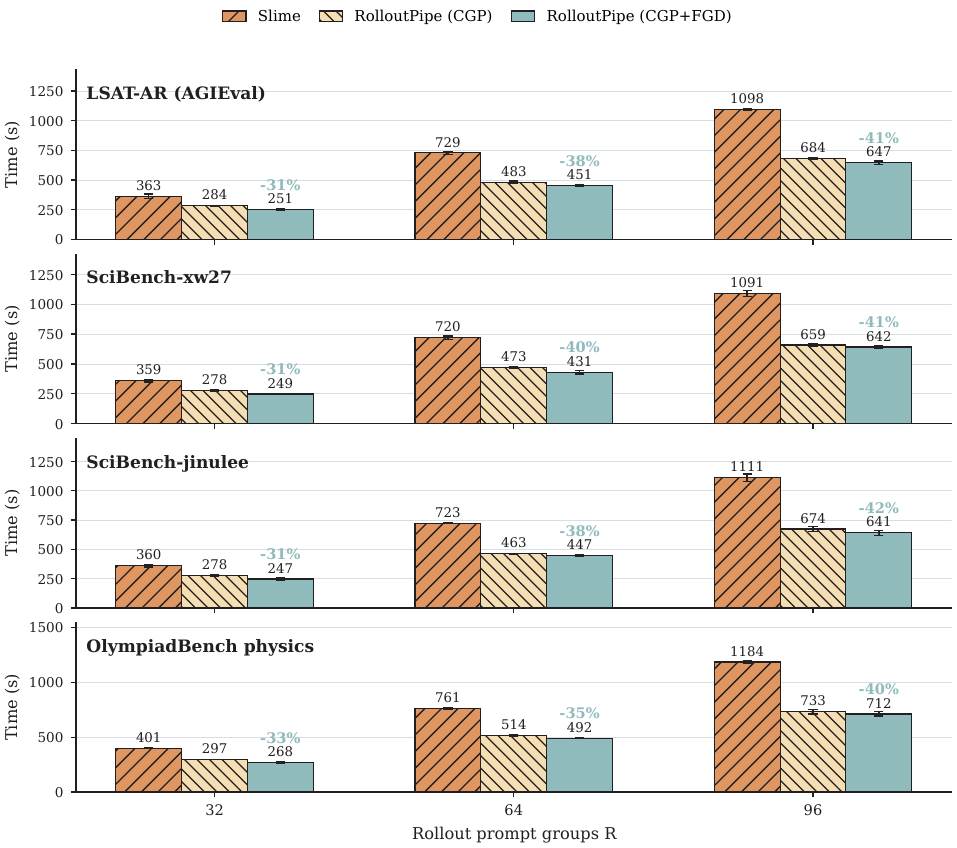}
\caption{Rollout-to-train-end time.
Each bar shows four-round mean with error bars indicating sample standard deviation,
and orange numbers show the percentage CGP+FGD shortens over Slime.}
\label{fig:rolloutpipe-main-window}
\end{figure}

To evaluate the overall speedup of RolloutPipe,
we measure the rollout-to-train-end main time under Slime, CGP, and CGP+FGD
across four workloads and three group counts.
As shown in Figure~\ref{fig:rolloutpipe-main-window},
CGP+FGD shortens the main time by 30.7\%--42.3\% over Slime across all twelve settings.
The reduction is consistent across LSAT-AR, Sci-XW, Sci-JL, and OlyPhys,
and grows with \(R\), from 30.7\%--32.9\% at \(R=32\) to 39.8\%--42.3\% at \(R=96\).
The reason is that a larger rollout leaves more trainable complete groups
trapped behind the rollout-completion barrier in Slime,
so overlapping their training with the remaining rollout reclaims more wait time.

To validate that CGP is the dominant source of the reduction,
we decompose the total gain into the CGP and FGD shares.
As shown in Figure~\ref{fig:rolloutpipe-main-window},
CGP accounts for 71\%--96\% of the total reduction, and its share grows with \(R\),
from 71\%--79\% at \(R=32\) to 92\%--96\% at \(R=96\).
In the serial mode, early-completed groups must wait for the last group to finish rollout
before entering training.
CGP removes this wait, and the larger the \(R\), the longer the removed wait,
which is why the CGP share rises with \(R\).

To measure how much FGD adds on top of CGP,
we compare CGP+FGD against CGP alone.
As shown in Figure~\ref{fig:rolloutpipe-main-window},
CGP+FGD further reduces the main time by 2.5\%--11.4\% over CGP at every \(R\).
The reason is that FGD prioritizes frontier groups at the serving-admission stage,
so completed groups flow into the trainer FIFO at steadier intervals
and reduce the idle periods where the trainer waits for the next \(U\)-group batch.

To demonstrate this supply-stabilizing effect,
we inspect dispatch trace and round completion times.
At \(R=96\), first legal dispatch drops from 509--543s in Slime
to 46--90s with CGP and 52--61s with CGP+FGD.
The CGP+FGD dispatch range (52--61s) is tighter than CGP (46--90s),
indicating frontier groups arrive more concentrated.
In Sci-XW at \(R=96\), full main time is 642s under CGP+FGD versus 659s under CGP,
and the gap comes from steadier group supply leaving the trainer waiting less mid-round.

\subsection{Training Workload and Wait Time Analysis}

\begin{figure}[!htbp]
\centering
\includegraphics[width=0.98\textwidth]{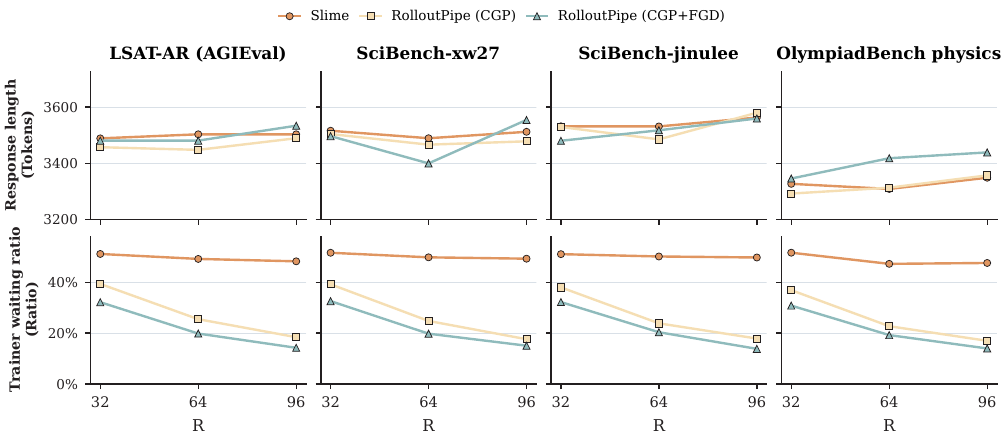}
\caption{Response length and trainer waiting ratio across workloads.
Each point is the four-round mean, and the response-length row uses a local zoom.
Columns correspond to workloads, and \(R\) is the number of rollout prompt groups.}
\label{fig:rolloutpipe-training-stability}
\end{figure}

To evaluate that the reduction does not come from doing less training work,
we report response length and trainer compute time.
As shown in Figure~\ref{fig:rolloutpipe-training-stability} and Table~\ref{tab:trainer-compute},
response lengths stay in the same 3.3k--3.6k token range across configurations,
and trainer compute time is nearly identical across the three configurations at the same \(R\).
The reason is that all configurations form the same number of \(U\)-group logical batches
using the same Megatron microbatching, objective, and optimizer path.
The table reports active trainer compute, not total elapsed wall time,
so the nearly identical values in each row directly confirm that
RolloutPipe changes the placement of work, not the amount of work.

To validate that the reduction comes from trainer waiting,
we report the trainer waiting ratio.
As shown in Figure~\ref{fig:rolloutpipe-training-stability},
Slime submits complete groups only after the rollout completes,
resulting in a waiting ratio of 47\%--52\%,
while CGP+FGD reduces this to 14\%--33\% across the twelve reported points,
reaching 14.2\%, 15.0\%, 13.8\%, and 13.9\% at \(R=96\).
The reason is that RolloutPipe moves the same trainer computation earlier
on the wall-clock timeline while the remaining rollout is still running.

{\renewcommand{\arraystretch}{0.95}%
\setlength{\intextsep}{6pt}%
\begin{table}[t]
\centering
\caption{Trainer compute time per workload and \(R\) (seconds).}
\label{tab:trainer-compute}
\small
\resizebox{\textwidth}{!}{%
\begin{tabular}{lccc@{\hspace{1.5em}}ccc}
\toprule
 & \multicolumn{3}{c}{LSAT-AR} & \multicolumn{3}{c}{Sci-XW} \\
\cmidrule(lr){2-4} \cmidrule(lr){5-7}
\(R\) & Slime & RolloutPipe (CGP) & RolloutPipe (CGP+FGD) & Slime & RolloutPipe (CGP) & RolloutPipe (CGP+FGD) \\
\midrule
 32 & 195.1 & 194.7 & 195.9 & 191.8 & 191.7 & 192.5 \\
 64 & 388.9 & 387.6 & 391.9 & 379.9 & 383.4 & 375.9 \\
 96 & 587.4 & 589.0 & 586.4 & 572.2 & 573.0 & 577.3 \\
\midrule
 & \multicolumn{3}{c}{Sci-JL} & \multicolumn{3}{c}{OlyPhys} \\
\cmidrule(lr){2-4} \cmidrule(lr){5-7}
\(R\) & Slime & RolloutPipe (CGP) & RolloutPipe (CGP+FGD) & Slime & RolloutPipe (CGP) & RolloutPipe (CGP+FGD) \\
\midrule
 32 & 194.1 & 195.2 & 192.5 & 211.5 & 210.6 & 211.1 \\
 64 & 384.6 & 380.7 & 385.7 & 420.7 & 425.7 & 426.9 \\
 96 & 584.5 & 584.1 & 584.1 & 640.0 & 640.0 & 644.4 \\
\bottomrule
\end{tabular}%
}
\end{table}}

To ensure that pipelining overlaps training with rollout rather than changing workload,
we compare trainer compute time and response length at the same \(R\).
Table~\ref{tab:trainer-compute} shows no significant difference in trainer compute time,
indicating unchanged training computation;
response length is consistent, indicating comparable rollout workload.
On OlympiadBench physics, CGP+FGD has longer response length yet shorter rollout-to-train-end window than Slime, proving speedup comes from overlap not reduced computation.
Unlike asynchronous pipelines, RolloutPipe never uses stale data: it pipelines complete groups generated by fixed rollout weights, preserving on-policy correctness.
RolloutPipe preserves on-policy semantics: all \(R\) groups share fixed rollout weights, and optimizer-step boundary follows Slime's \(U\)-group logical updates.
Since CGP and FGD change only group dispatch timing and order—not samples, advantages, or gradient computation—training dynamics and convergence behavior remain consistent with Slime.

\section{Related Work}

Related work spans three areas.
RLHF and RLVR infrastructure, asynchronous and pipelined RL training systems,
and rollout runtime scheduling.
RolloutPipe targets the synchronous on-policy path represented by Slime.
It pipelines trainable complete groups generated by the same rollout weights.

\textbf{RLHF infrastructure and scheduling.}
HybridFlow, OpenRLHF, and RLHFuse provide infrastructure for rollout,
reward, training, and parameter synchronization~\cite{hybridflow,openrlhf,rlhfuse}.
They map post-training workflows onto stage orchestration and resource allocation.
Megatron-LM and Ray support large-model training and execution~\cite{megatronlm,ray};
serving systems like SGLang manage scheduling at request and token granularity~\cite{sglang}.
RolloutPipe changes rollout-to-trainer handoff granularity:
FGD directs scheduling toward frontier groups forming the next training batch,
CGP delivering them after materialization.

\textbf{Asynchronous and pipelined RL training systems.}
AReaL and AsyncFlow reduce
rollout-training bubbles by overlapping generation and training
using asynchronous workflows~\cite{areal,asyncflow}.
Their overlap is organized around generation streams, task batches, or asynchronous weight refresh,
which trades off data freshness for pipeline overlap, thus not on-policy.
RolloutPipe instead uses trainable complete groups from the same fixed-weight rollout,
forming logical updates from \(U=B/K\) complete groups and
pipelining complete-group dispatch within the same on-policy round under trainer FIFO semantics.

\textbf{System architecture and scheduling in cognitive computing.}
Cognitive computing systems face related challenges in staging generation and consumption.
Multimodal RAG systems adopt offline preparation and online retrieval pipelines~\cite{iccc2025multimodalrag},
adaptive ranking algorithms prioritize items by real-time signals~\cite{iccc2024hotranking},
and multi-agent systems coordinate distributed workers~\cite{iccc2025ecommerce}.
These works stage or prioritize independent units to optimize global throughput or latency.
RolloutPipe instead couples rollout and training under on-policy semantics with all groups sharing fixed weights and each GRPO group trainable only after materialization, so pipelining must respect the group boundary.
FGD targets only the frontier groups forming the next training batch, prioritizing their stable supply to the trainer rather than optimizing global throughput or per-request latency.

\section{Conclusion}

In synchronous on-policy GRPO training with decoupled resources,
RolloutPipe pipelines complete groups out of an ongoing fixed-weight rollout.
CGP delivers trainable groups early, FGD applies frontier-first admission,
and the training side consumes \(U\)-group logical batches,
overlapping rollout and training under on-policy semantics.
Across twelve result points on multiple workloads,
CGP+FGD shortens the main time by 30.7\%--42.3\%
and lowers the trainer waiting ratio by 37\%--76\% relative to Slime.
One direction remains for future work.
Since RLVR embodies a perceive-reason-update loop of cognitive computing,
RolloutPipe's pipelining can generalize to other cognitive systems where generation and learning are disaggregated and must be overlapped.

\begin{credits}
\subsubsection{\ackname}
	This work is supported by the Guangdong Science and Technology Cooperation Project (No. 2025A0505020065), Guangdong Basic and Applied Basic Research Foundation (No. 2024A1515010251), Key Research and Development and Technology Transfer Program of Inner Mongolia Autonomous Region (2025YFHH0110) and Shenzhen Basic Research Program (No. JCYJ20240813155810014).
\end{credits}

\renewcommand{\doi}[1]{}
\renewcommand{\url}[1]{}
\renewcommand{\doi}[1]{}
\bibliographystyle{splncs04}
\bibliography{references}

\end{document}